\documentclass[sigconf, authorversion,nonacm]{acmart}

\usepackage[nameinlink]{cleveref}
\usepackage{enumitem}
\usepackage{float}
\usepackage{amsmath}
\usepackage{makecell}
\usepackage{algorithm}
\usepackage{subfig}
\usepackage{stmaryrd}
\usepackage{xspace}
\usepackage{comment}
\usepackage[noend]{algpseudocode}
\newtheorem{example}{Example}
\usepackage{tablefootnote}
\usepackage{soul}
\usepackage{balance}
\usepackage[a-1b]{pdfx}
\AtBeginDocument{%
  \providecommand\BibTeX{{%
    \normalfont B\kern-0.5em{\scshape i\kern-0.25em b}\kern-0.8em\TeX}}}

\setcopyright{none}

\begin{document}
\title{Discovering Graph Generating Dependencies for Property Graph Profiling}

\author{Larissa C. Shimomura}
\email{larissa.capobianco-shimomura@ipvs.uni-stuttgart.de}
\orcid{0009-0008-4679-7656}
\affiliation{%
  \institution{IPVS - University of Stuttgart}
  \city{Stuttgart}
  \country{Germany}
}

\author{Nikolay Yakovets}
\email{n.yakovets@tue.nl}
\orcid{0000-0002-1488-1414}
\affiliation{%
  \institution{Eindhoven University of Technology}
  \city{Eindhoven}
 \country{Netherlands}
}

\author{George Fletcher}
\email{g.h.l.fletcher@tue.nl}
\orcid{0000-0003-2111-6769}
\affiliation{%
  \institution{Eindhoven University of Technology}
  \city{Eindhoven}
  \country{Netherlands}
}

\begin{abstract}
Knowledge graphs have soared in popularity by supporting different types of applications and domains. In this context, the property graph data model has become an emerging standard in industry and academia. With its widespread use, there is also an increasing interest in investigating constraints for property graph data and their applications in data profiling.
Graph Generating Dependencies (GGDs) are a class of property graph data dependencies that can express constraints on topology and properties of nodes and edges of the graph, making them a suitable candidate to expose an overview of the property graph to the user (profile graph data). 
However, GGDs can be difficult to set manually. To solve this issue, we propose a framework for discovering GGDs automatically from the property graph to profile graph data. Our framework has three main steps: (1) \textit{pre-processing}, (2) \textit{candidate generation}, and, (3) \textit{GGD extraction}. 
Our results show that the discovered set of GGDs can give an overview of the input graph, including schema-level information between the graph patterns and attributes.
\end{abstract}

\ccsdesc[500]{Information systems~Graph-based database models}

\keywords{Graph Data Dependencies; Graph Generating Dependencies; Discovery of Dependencies; Property Graph; Data Profiling}

\maketitle

\section{Introduction}

Data dependencies can be understood as constraints imposed on a dataset that can be used to aid in general understanding of the data ~\cite{Liu2012,2018Abedjan}. 
When dealing with graph data, information about topology and attributes of the nodes/edges is essential. The property graph data model (see formal definition in~\cite{Bonifati2018}) is an emerging standard in the industry. Consequently, there is an interest in developing and researching data dependencies as tools for applications such as data profiling.
 
Graph Generating Dependencies (GGDs)~\cite{Shimomura2020} is a class of dependencies proposed for property graphs that can express that for every homomorphic match of a source graph pattern, there should exist a homomorphic match of a (possibly) different target graph pattern. Informally, GGDs can express constraints between two (possibly) different graph patterns and the similarity of the properties of its nodes and edges (see formal definition in \Cref{sec:preliminaries}).

GGDs can fully capture tuple-generating dependencies (tgds) and can enforce the existence of a node, edge, or graph pattern compared to previously proposed dependencies for property graph, e.g., Graph Entity Dependencies (GEDs)~\cite{Fan2019} and Graph Differential Dependencies (GDDs)~\cite{Kwashie2019}, that generalize equality-generating dependencies that mean, enforce equality (or similarity in case of the GDDs) on the property values of a graph pattern.

\begin{figure}[t]
    \centering
    \includegraphics[width=0.7\linewidth]{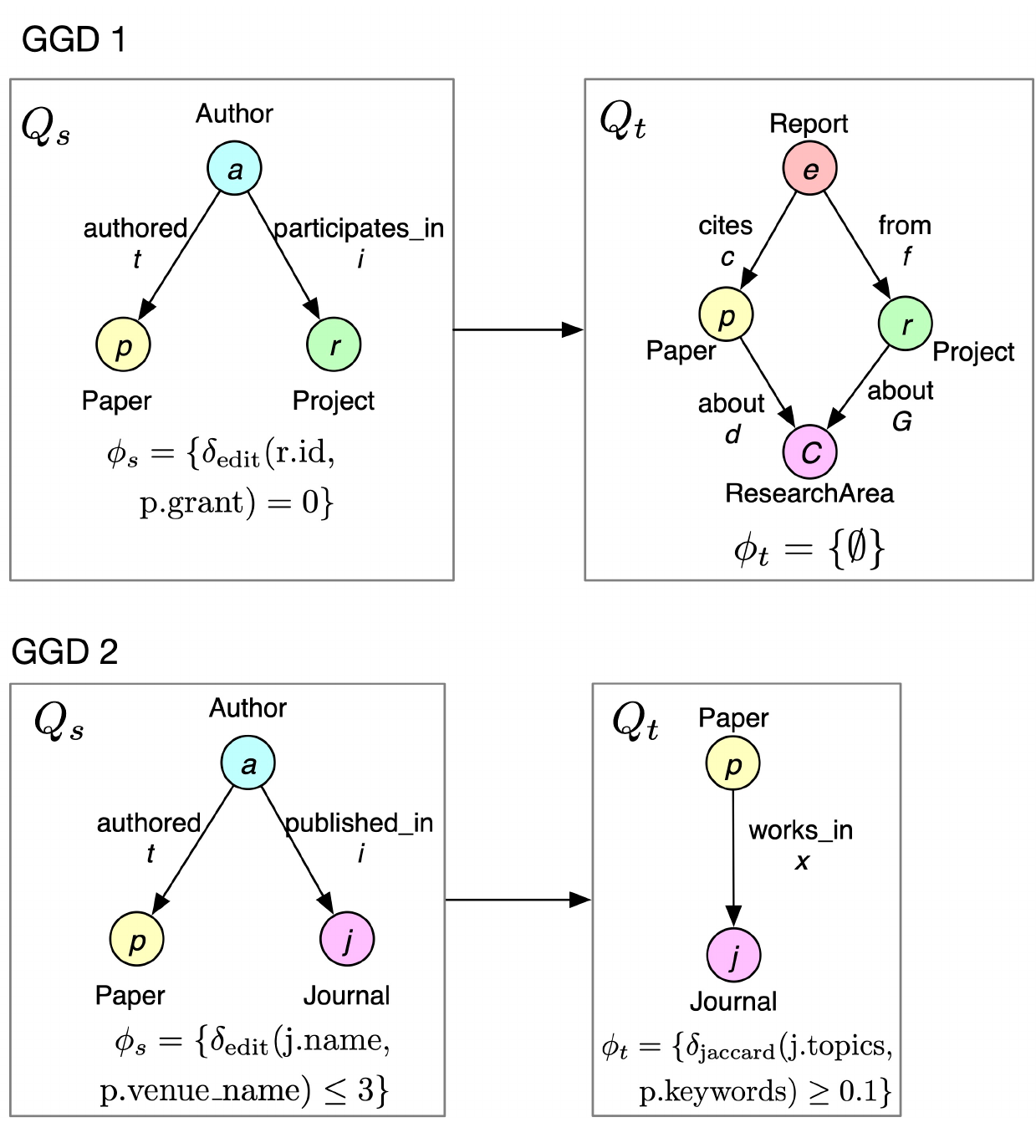}
    \caption{Example GGD}
    \label{fig:exampleGGDIntro}
\end{figure}

GGDs introduce a new expressive power needed to capture constraints that enforce the existence of a node, edge, or a new graph pattern. This expressive power is needed in property graphs as relationships are first-class citizens in this data model, and correlation of different graph patterns can naturally arise. 
Other novelties of GGDs include~\cite{Shimomura2020,Shimomura2022}: (i) can express constraints on edge attributes (not considered in GEDs) and (2) can express constraints on node and edge attributes according to their similarity.
Consider the GGDs in \autoref{fig:exampleGGDIntro} in the context of a publication network. 

\begin{example}
    GGD 1 expresses a constraint on the existence of a relation to another vertex. For every author who has participated in a project and authored a paper in which the paper funding grant number is equal to the project number, there should exist a Report node related to the Project that cites the Paper, and they should be about the same ResearchArea node. 
\end{example}

\begin{example}
    GGD 2 expresses a constraint on the attributes of a graph pattern. For every Paper written by an Author that has an edge to a Journal indicating publication and where the paper venue name and the journal name are similar, there should exist an edge connecting the node Paper to the Journal labeled ``appeared\_in'' and the journal topics are similar to the paper keywords.  
\end{example}

GGD 1 enforces the existence of nodes and edges in the graph, giving information about how connected the nodes labeled Author, Paper, Project, Report, and Research Area are. GGD 2 enforces a similarity of the values and gives information about how related a Paper connected to a Journal are in terms of its attributes, as whenever this dependency holds it means that the paper and journal topics, venue name, and name are similar. 

These GGDs help data scientists understand the underlying relations between the nodes/edges and their attributes in the property graph and can be used by data engineers to manage the graph.
To the best of our knowledge, GGD is the only formalism that can fully capture tgds and similarity constraints for property graphs. 

However, manually defining GGDs is time consuming and requires the knowledge of a data expert. In this paper, we propose GGDMiner, a framework for automatic discovery of \emph{approximate} Graph Generating Dependencies (GGDs), GGDs that hold in most of the data with a certain degree of inconsistency. In particular, we focus on the task of data profiling by discovering a set of GGDs that can give an overview of the graph data. 

The GGDMiner framework consists of three main steps: (1) Pre-processing - builds indexes to assist in the process of discovering the similarity constraints in the next step. (2) Candidate Generation - mines graph patterns and differential constraints that might be a source or target of a GGD. 
(3) GGD Extraction - verifies which candidates generated in the previous step can be paired to form a GGD that is interesting for describing the graph data. 
We give more details about the general framework and each step in \Cref{sec:ggdminer}. 

To scale to more complex graphs, in GGDMiner we use a factorized representation of each candidate called Answer Graph~\cite{answergraph}. The answer graph was previously proposed for evaluating graph pattern queries and to the best of our knowledge, our work is the first time the Answer Graph method is used in the scenario of discovery of constraints. The factorized representation of Answer Graph allows us to operate on the matches of each graph pattern without defactorizing it to a table-like representation.
 
The goal of GGDMiner is to provide a baseline solution for the discovery of GGDs with a clear step-by-step approach leveraging state of the art pattern mining algorithms in combination with our novel approach based on Answer Graph. 
The rest of the Paper is structured as follows. In \Cref{sec:related} and \Cref{sec:preliminaries}, we present related work and an overview of the GGDs definition. \Cref{sec:dataprofiling} and \Cref{sec:ggdminer} present the problem definition of discovering GGDs and the GGDMiner framework, followed by experiments and results in \Cref{sec:experiments}.

\section{Related Work}
\label{sec:related}

We place this work in the context of discovery algorithms for relational and graph data dependencies and frequent subgraph mining.

\paragraph{Relational Data Dependency Discovery} - 
Functional dependencies (FDs) are the most used type of dependencies used in relational data. According to \cite{Liu2012}, there are mainly two approaches for the discovery of FDs: (1) top-down algorithms, also called column-based algorithms, and (2) bottom-up methods, also called row-based algorithms. 
These two types of approaches were also used for the discovery of other types of dependencies, such as Conditional Functional Dependencies (CFDs)~\cite{Fan2011}, Differential Dependencies (DDs)~\cite{Song2011}, Matching Dependencies (MDs)~\cite{Schiemer2020} etc.
A common way to represent the dependency candidates is through a lattice. How to build and traverse the lattice and the pruning techniques can vary according to the type of dependency (see \cite{Liu2012,Falco2019} for details).

\paragraph{Graph Data Dependency Discovery} - Graph dependencies also add the challenge of discovering information about the topology. 
In the paper ~\cite{Kwashie2019}, the authors propose a lattice-based algorithm for discovering Graph Differential Dependencies (GDDs) for entity resolution. 
In ~\cite{Fan2020}, the authors proposed a parallel algorithm for discovering Graph Functional Dependencies (GFDs). Starting from a single node pattern, this algorithm is based on two main processes: (1) vertically spawning the search space to extend the graph pattern and (2) horizontally spawning the search space to discover the dependency literals of the GFDs.
Besides graph data dependency discovery algorithms, algorithms for mining graph association rules such as the discovery of GPARs (Graph Pattern Association Rules)~\cite{Fan2022} and AMIE+~\cite{Amie2015} have also been proposed.
Proposed algorithms of the literature use either a lattice or a similar strategy to explore the search space~\cite{alipourlangouri2022,fastageds}. In GGDMiner, we propose a similar strategy in the Candidate Generation step. However, more than this step is needed to discover GGDs. 
The discovery of GGDs is more challenging compared to other classes of dependencies for property graphs that are defined over a single graph pattern, such as GEDs or GFDs, as for GGDs, we also need to discover which graph patterns and attribute values are associated with each other.
In GGDMiner, the lattice structure is used only to identify which graph patterns and constraints are frequent. To discover GGDs, we introduce the candidate index and the use of Answer Graph~\cite{answergraph} to identify which of the candidates co-occur. The candidate index and the Answer Graph are a novelty of GGDMiner.

\paragraph{Frequent subgraph mining} - Frequent subgraph mining (FSM) refers to the task of finding all isomorphic subgraphs that occur more than a designated number of times in a graph~\cite{Jiang2013}. 
A well-known FSM algorithm is gSpan~\cite{Yan2022}. The main idea of gSpan is to map each subgraph to a canonical DFS code. The DFS codes are used to check if two subgraphs are isomorphic.
The DFS codes were largely used in other algorithms, including state-of-the-art algorithms such as GRAMI~\cite{Elseidy2014}. 
GRAMI models the frequency evaluation of each subgraph as a constraint satisfaction problem.
GRAMI has been extended to support variations of the FSM problem~\cite{Elseidy2014,Nguyen2021}.

\section{GGDs - Definition Overview}
\label{sec:preliminaries}

A Graph Generating Dependency~\cite{Shimomura2020} is a dependency of the type $Q_s[\overline{x}]\phi_s \rightarrow Q_t[\overline{x},\overline{y}]\phi_t$ in which $Q_s[\overline{x}]$ is called the source graph pattern in which $\overline{x}$ is the set of variables (nodes and edges) in the graph pattern and $\phi_s$ is the set of differential constraints over $\overline{x}$ and $Q_t[\overline{x},\overline{y}]$ is the target graph pattern which is a graph pattern that can contain variables from the source graph pattern ($\overline{x}$) and additional variables ($\overline{y}$) and $\phi_t$ is a set of target differential constraints over the variables $\overline{x},\overline{y}$.

The differential constraints of the set $\phi_s$ and $\phi_t$ can be of the form: (1) $\delta(x.A, c) \le t_A$, (2) $\delta(x.A_1, x.A_2) \le t_{A_1,A_2}$ or (3) $x = x'$ on which $\delta$ is a user-defined (dis)similarity function, $x.A$ refers to property $A$ of the variable $x$, $c$ is a constant and $t_A$ is a user-defined threshold. The first two types of constraints compare a property value to a constant or another node/edge property according to the threshold $t_A$ and (dis)similarity function $\delta$. The third type checks if variables $x$ and $x'$ refer to the same node or edge in the graph. 

Given a property graph $G$, we say a GGD $\sigma$ is satisfied if for all matches $h_s[\overline{x}]$ of the source graph pattern $Q_s$ in $G$ which satisfies the source differential constraints $\phi_s$, there should exist a match $h_t[\overline{x},\overline{y}]$  in $G$ of the target graph pattern $Q_t$ which satisfies the target differential constraints $\phi_t$. Due to space limitation, see \cite{Shimomura2020,Shimomura2022} for more details about GGDs.

In this work, we are interested in the discovery of \emph{Extension GGDs}, a GGD in which at least one variable is explicitly part of both source and target graph patterns. Informally, it means that the target is an extension of the source. The GGDs of \autoref{fig:exampleGGDIntro} are extension GGDs. For example, the nodes ``p'' (Paper) and ``r''(Project) in GGD1 appear in source and target graph patterns.

\section{GGD Discovery - Problem Definition}
\label{sec:dataprofiling}

GGDMiner aims to discover a set of GGDs for profiling property graphs, or, in other words, mining a set of GGDs that can give the user an overview of the graph data.
To have an overview of the graph data, it is interesting to understand what kind of relations (graph patterns and properties) frequently appear in the graph. Therefore, if something happens frequently in a dataset, it should be considered important to describe the dataset. This assumption was also used in previous discovery algorithms~\cite{alipourlangouri2022,Fan2020}.

We are interested in discovering GGDs in which both source and target components happen frequently in the graph (defined as support of source/target) and frequently co-occur in the graph (defined as confidence of a GGD). In the following, we present our definitions of support and confidence for GGDs.

\paragraph{Support} 
Given a GGD $Q_s[\overline{x}]\phi_s \rightarrow Q_t[\overline{x},\overline{y}]\phi_t$, we define support of the source, denoted as $supp(Q_s[\overline{x}]\phi_s)$ (resp. support of the target, $supp(Q_t[\overline{x}\overline{y}]\phi_t)$), as the number of source matches in the graph $G$ that satisfies the source constraints, denoted as $|h_s[\overline{x}] \models \phi_s|$ (resp. on the target, $|h_t[\overline{x}\overline{y}] \models \phi_t|$). 

\paragraph{Confidence} Confidence quantifies how much a GGD holds in the data. According to GGD semantics, a GGD is validated if, for all matches of the source there exists a match of the target. Considering $\alpha = Q_\alpha[\overline{x}]\phi_\alpha$ and $\beta = Q_\beta[\overline{y}]\phi_\beta$ and there exists at least one variable $a \in \overline{x}$ in $\alpha$ which can match to the same nodes/edges to a variable $b \in \overline{y}$ in $\beta$ then the confidence of the GGD $\alpha \rightarrow \beta$: \[conf(\alpha \rightarrow \beta) = \frac{|Validated(\alpha \rightarrow \beta)|}{supp(\alpha)},\] in which $|Validated(\alpha \rightarrow \beta)|$ is the number of matches of $\alpha$ in which the possible GGD $\alpha \rightarrow \beta$ is validated (satisfied) and $supp(\alpha)$ is the support of the source $\alpha$.

Complementary to frequency, we are interested in discovering a set of GGDs that maximizes how much we can describe the graph data. 
\paragraph{Coverage} - Coverage is a measure used in previous works in the literature to define how much of the data a particular dependency can give information about the data.
Since the main goal is to give the user an overview, we consider a set of GGDs that can describe the graph data as a set of GGDs that can give information about the biggest number of nodes and edges in the graph. Following the semantics of the GGDs, we are mainly interested in how much information the source side can give about the data. We define the coverage of a set of GGDs $\Sigma_G$ as:

\[coverage(\Sigma_G) = \frac{|\cup_{\sigma \in \Sigma_G}O(\sigma)|}{|G|},\] in which $O(\sigma)$ is the set of matching nodes and edges of the source side of the GGD and $|G|$ is the total number of nodes and edges in the input graph $G$.

Data dependency discovery algorithms from the literature focus on discovering a minimal set of non-redundant dependencies, also called minimal cover. However, the problem of identifying redundant GGDs in a set has high computational complexity (see formal results for Implication in \cite{Shimomura2022}). To minimize the number of similar or redundant GGDs, we introduce the decision boundary and candidate similarity to avoid similar GGDs in the result set.

\paragraph{Decision Boundary} -  The decision boundary is a tuple $<\upsilon,\kappa>$ defined for each attribute data type in the graph data, for example, strings, numbers, sets, dates, etc. 
The $\upsilon$ refers to the minimal threshold (dissimilarity) value and the $\kappa$ is the minimal difference between two threshold values. We say that a set $\phi$ of differential constraints on the same attribute and same constant values respect a decision boundary if the smallest discovered threshold $t$ of the set is bigger than $\upsilon$ and the smallest difference between all thresholds of the set $\phi$ is bigger than $\kappa$. The decision boundary is used to avoid the discovery of very similar differential constraints.

\begin{example}
Consider $\alpha = Q_\alpha[\overline{x}]\phi_\alpha$ and two differential constraints discovered over the string attribute name $\phi_\alpha = \{\delta_{\text{name}}(x.name,$ $ x.surname)  \le 3.0$ , $\delta_{\text{name}}(x.name, x.surname) \le 1.0\}$, and a decision boundary defined for string attributes as $<1.0, 2.0>$. The set of constraints $\phi_\alpha$ respects the decision boundary, as all thresholds of the constraints in $\phi_\alpha$ are at least $1.0$ and the smallest difference between its threshold is $2.0$ ($3.0 - 1.0 = 2$).
\end{example}

\paragraph{Candidate Similarity} - Considering $\alpha = Q_\alpha[\overline{x}]\phi_\alpha$ and $\beta = Q_\beta[\overline{y}]\phi_\beta$, we measure the similarity between $\alpha$ and $\beta$ according to 2 aspects: (i) Graph pattern denoted as $\delta_Q(Q_\alpha,Q_\beta)$, (ii) Differential constraints, denoted as $\delta_\phi(\phi_alpha, \phi_\beta)$. Given these two aspects, the overall similarity $\Delta$ between $\alpha$ and $\beta$ is defined as: \[\Delta(\alpha,\beta) = 
\frac{\delta_Q(Q_\alpha, Q_\beta) + \delta_\phi(\phi_\alpha, \phi_\beta)}{2},
\] in which $\delta_Q(Q_\alpha, Q_\beta)$ is the number of common edges (same edge label, source node label, and target node label) and $\delta_\phi(\phi_\alpha, \phi_\beta)$ is the number of differential constraints that refer to the same attribute of the same node/edge label.

Considering the coverage measure, given an input property graph $G$ and assuming that we have information about the attributes and the domain that each node/edge label has (see \autoref{fig:schema}), we define the problem of discovering GGDs for property graph profiling as follows.

\textit{Input:} Given an input property graph $G$ and its set of labels and attributes, support threshold $\tau$, a set $T$ of decision boundary for each domain of values, a confidence threshold $\epsilon$, a similarity value $\theta$ and a maximum number of edges $k$.

\textit{Output:} A set of Extension GGDs $\Sigma_G$ with maximum coverage in which each GGD in $\Sigma_G$ has confidence bigger than $\epsilon$. And, the source and target of each GGD that has support is bigger than $\tau$.

Besides the support $\tau$ and confidence $\epsilon$ values, we also introduce the parameter $k$ and the parameter $\theta$, in which $k$ refers to the maximum number of edges each graph pattern (source or target) in a GGD can have and, $\theta$ is a similarity threshold which defines how similar a GGD can be from another GGD from the discovered set $\Sigma_G$.
Such parameters are introduced because GGDMiner aims to give the user an initial overview of the property graph, and very complex and similar graph patterns might be hard to understand at first glance.

An important step in discovery algorithms is candidate generation. 
As discussed in previous works~\cite{Liu2012,2018Abedjan}, the number of candidate dependencies to be considered can be exponential to the number of attributes of the data. 
By definition, each differential constraint on a GGD can be according to a user-defined distance measure. To reduce the scope of the candidate generation, we fix a distance measure for each attribute domain. We use the edit distance for string values and the absolute difference for numerical values.

\section{GGDMiner Framework}
\label{sec:ggdminer}

\begin{figure}[t]
    \centering
        \includegraphics[width=0.8\linewidth]{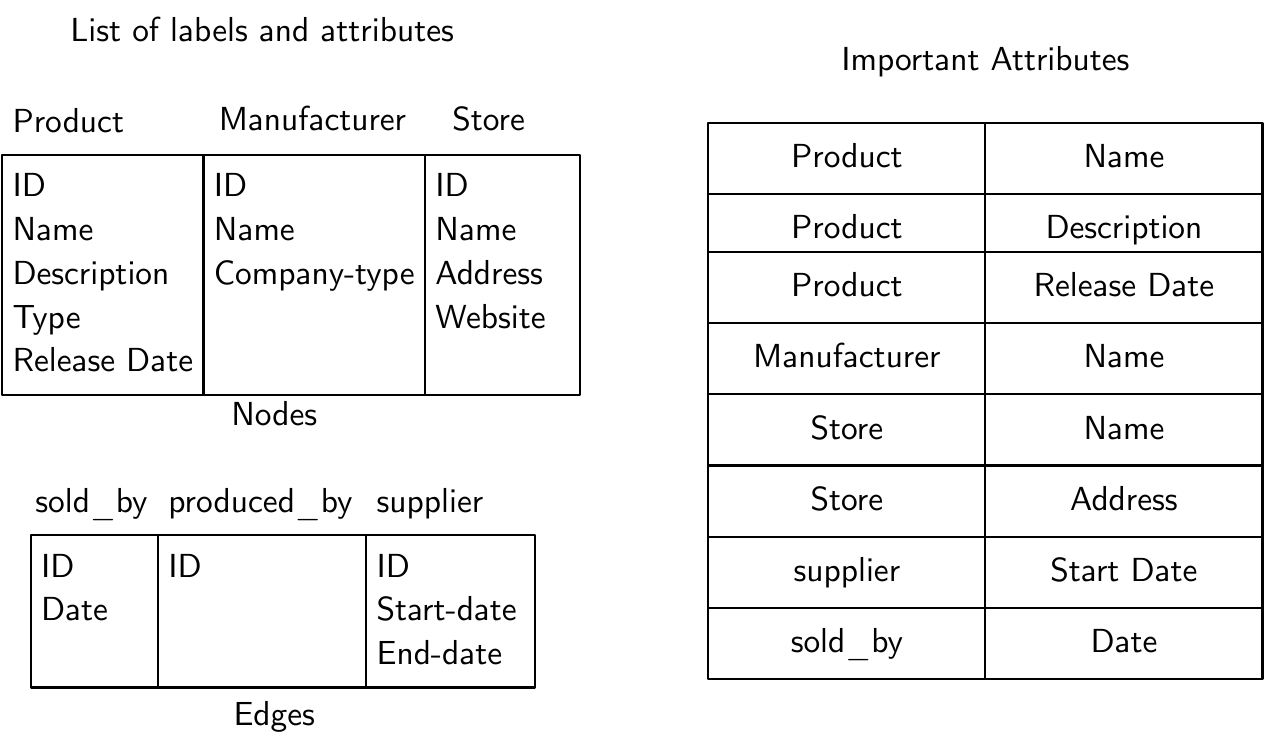}
    \caption{Example Graph Labels and Attributes and Important attributes selected}
    \label{fig:schema}
\end{figure}

As mentioned, the GGDMiner framework has three main steps: (1) Pre-processing, (2) Candidate Generation and (3) GGD Extraction. In this section, we give details of each one of the steps. We use a graph which contains the labels and attributes in \Cref{fig:schema} as a running example throughout the section. 

\begin{figure}[t]
    \centering
    \includegraphics[width=\linewidth]{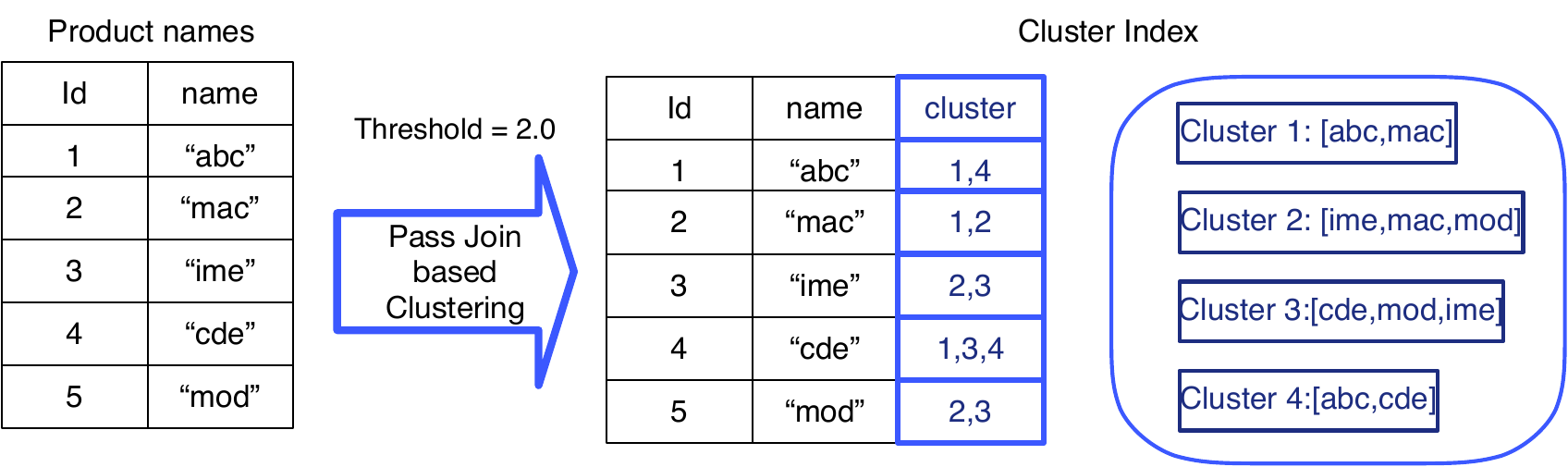}
    \caption{Similarity Clustering Index}
    \label{fig:simIndex}
\end{figure}%

\subsection{Pre-processing}

In the pre-processing step, the main tasks are: (1) Identifying which are possible extensions and important attributes to consider during the candidate generation and (2) Constructing similarity indexes for the important attributes that will be used in the discovery of differential constraints in the Candidate Generation step.

\paragraph{Selecting Attribute Pairs}
In this task, we select which are the attributes of each node/edge label that can possibly be part of a differential constraint of the type $\delta_\text{attr}(x.attr_1, y.attr_2) \le t_{x,y}$ or $\delta_\text{attr}(x.attr_1, c) \le t_{x,c}$, considering that $x$ and $y$ are variable of the same graph pattern with (possibly) different labels, $c$ is a constant value and $attr_1$ and $attr_2$ are attributes of the same domain. 
Based on the assumption that attributes with similar values should also have semantically similar names, we select attributes with semantic similarity above a user-defined threshold. 
The pairs and the attributes that are part of a pair with semantic similarity above a user-defined threshold are added to a list of important attributes. This list is used to reduce the number of candidates of differential constraints in the next step of the framework as only the pairs of attributes in this list are considered for mining differential constraints of the type $\delta_\text{attr}(x.attr_1, y.attr_2) \le t_{x,y}$, and only attributes that appear in any pair of the list are considered for mining differential constraints of the type $\delta_\text{attr}(x.attr_1, c) \le t_x$. \autoref{fig:schema} shows the important attributes list of our running example.

\paragraph{Similarity Indexes} To discover differential constraints of the type $\delta_\text{attr}(x.attr_1, \varrho) \le t_{x}$, we build similarity indexes according to the domain of $x.attr_1$ to speed up the discovery process of this type of constraints in the Candidate Generation step. 
For each one of the attributes that appear on the important attributes list, we build a structure that groups the attribute values according to their similarity and the threshold of its corresponding decision boundary domain. We call this structure a similarity cluster.
Given an attribute from the list and the minimal threshold $\upsilon$ defined for the domain of this attribute in the decision boundary set $T$, we first select all the values of this attribute in the graph. We then execute a string similarity join algorithm with threshold $\upsilon$ to get all pairs of values that are similar according to $\upsilon$. In our implementation, we use the pass join algorithm~\cite{passjoin} due to its low memory consumption in previous comparisons to other similarity join algorithms~\cite{Jiang2014}. Finally, we build the similarity cluster by grouping all the output pairs that are joined by each other. Each similarity cluster is stored temporarily to be used in the next step of the framework. \Cref{fig:simIndex} shows an example of this process given a set of values of the attribute ``name''. Observe that using a similarity join algorithm to build the clusters allows one attribute value to be part of multiple clusters. We explain how we use the similarity cluster with more details in \Cref{subsec:horizontalExpansion} 

\subsection{Candidate Generation}

The lattice is a data structure used in many dependency discovery algorithms to order and organize the dependency candidates. Similarly, in GGDMiner, a lattice structure is used to mine possible sources or target candidates for a GGD.
Each lattice node represents a candidate and corresponds to a graph pattern $Q[\overline{x}]$, a set of differential constraints $\phi$ and an answer graph $E$~\cite{answergraph}, a factorized representation of the matches of this graph pattern that satisfies the differential constraints $\phi$. 

\begin{figure}[t]
    \centering
    \includegraphics[width=0.8\linewidth]{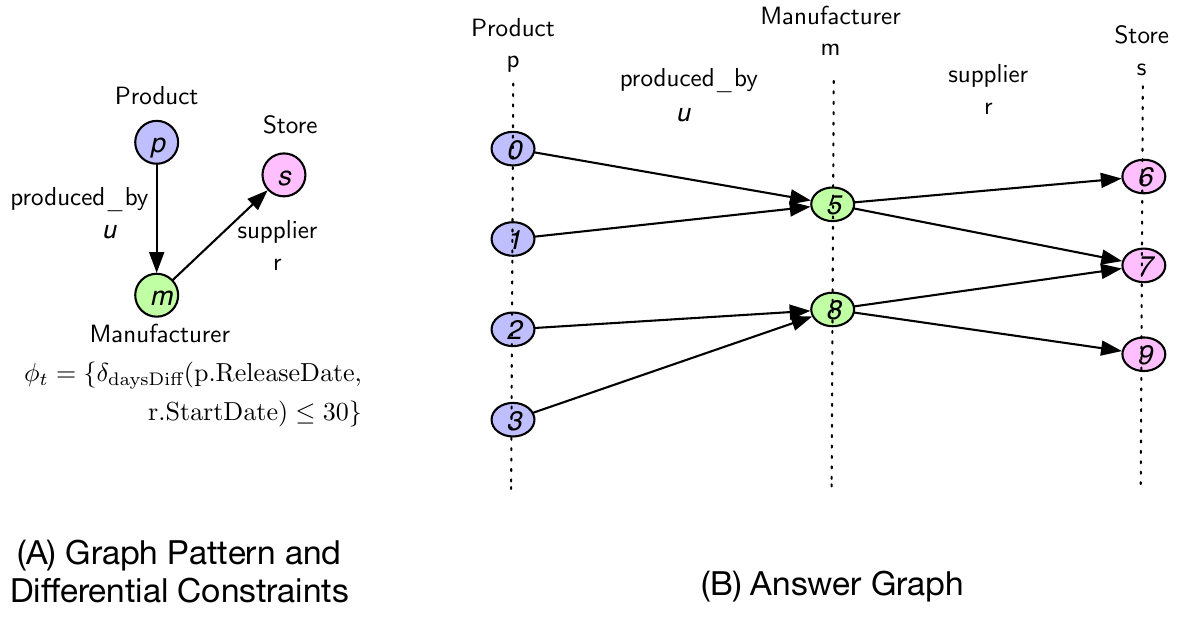}
    \caption{Answer Graph Example}
    \label{fig:answergraph}
\end{figure}

\paragraph{Answer graph} The answer graph is defined as a subset of the graph that suffices to compute the matches of a graph pattern~\cite{answergraph}. In GGDMiner, the answer graph is the intermediate representation of each lattice node's matching nodes and edges. The use of Answer Graph is one of the key components of GGDMiner, as it is a compact subgraph representation in which we can execute operations over the matching nodes and edges of each lattice node without needing to extract (defactorize) the full matches of the graph pattern in a table-like representation. Given a graph pattern, see the corresponding answer graph in \Cref{fig:answergraph}. 

\subsubsection{Lattice Construction - Discovery of Candidates}

\begin{figure}[t]
    \centering
    \includegraphics[width=\linewidth]{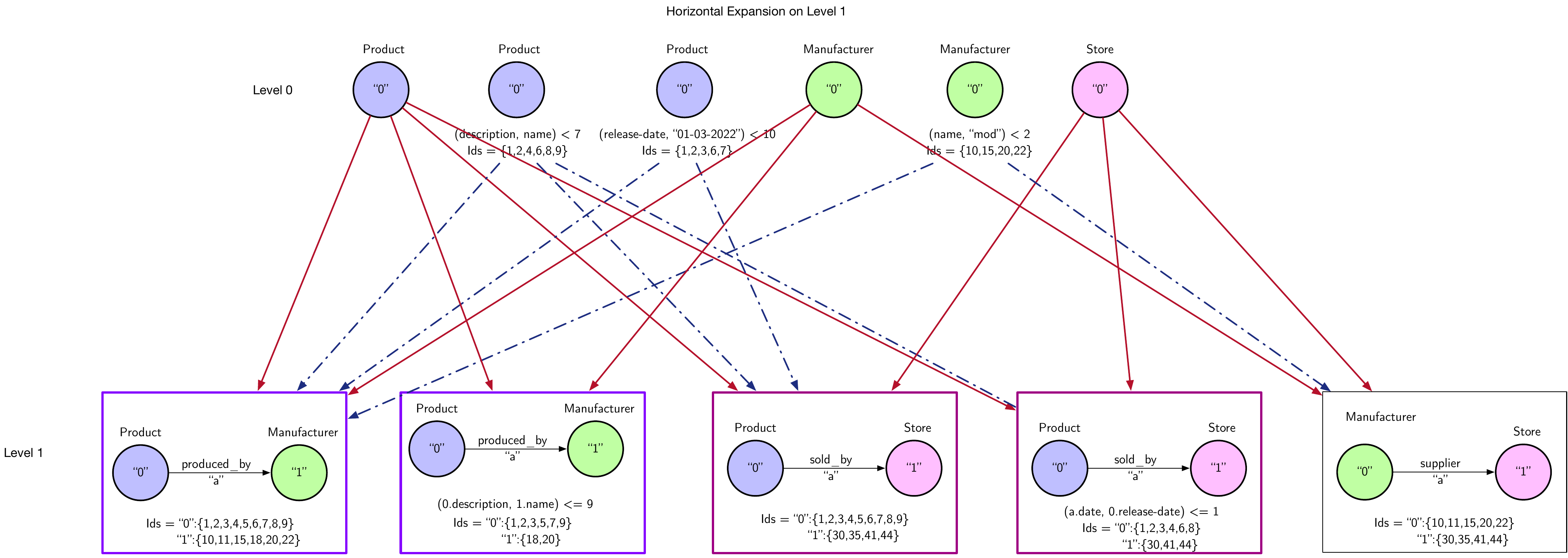}
    \caption{Level 1 of the lattice after vertical and horizontal expansion}
    \label{fig:secondlevel_2}
\end{figure}

As mentioned before, each lattice node refers to a graph pattern $Q[\overline{x}]$, a set of constraints $\phi$ and its matching nodes and edges represented by an answer graph $E$.
At each level of the lattice, we expand the graph pattern $Q[\overline{x}]$ by one edge, we call this process vertical extension. On the same level of the lattice, we expand the set of differential constraints considering the same graph pattern, we call this process horizontal expansion. 

\begin{algorithm}[t]
\footnotesize
    \caption{Vertical Expansion}
    \label{algo:verticalExpansion}
    \begin{algorithmic}[1] 
        \Procedure{VerticalExpansion}{Level Number level, Graph $G$, Support $\tau$, Lattice $L$, Candidate Index $C$, Lattice Node to extend \{ $Q$, $\phi$, $E$\}, Maximum number of edges $k$ , Frequent Edge Labels $L_e$}
        \State level $\gets$ level + 1
        \If{Number of edges in $Q \ge k$}
        \State \textbf{return}
        \EndIf
        \State fExtensions $\gets$  FSMSubgraphExtension($Q$, $G$, $\tau$, $L_e$) 
        \For{ each $Q_{ext} \in fExtensions$}
        \State $E_{ext}$ $\gets$ ExtendAnswerGraph($G$,$E$, $Q_{ext}$, $\emptyset$)
       \State L.addNode(level, $Q_{ext}$, $\emptyset$, $E_{ext}$)
       \State C.addCandidate($Q_{ext}$, $\emptyset$, $E_{ext}$)
        \State HorizontalExpansion(level, $\{Q_{ext}, \emptyset, E_{ext}\}$, $\tau$, L, C)
        \State VerticalExpansion(level, $G$, $\tau$, L, C, $\{Q_{ext}, \emptyset, E_{ext}\}$, $k$, $L_e$)
        \EndFor 
        \EndProcedure
    \end{algorithmic}
\end{algorithm}

We initialize the lattice by adding nodes with a single vertex graph pattern and an empty set of differential constraints for each vertex label of the graph $G$, which has support bigger than $\tau$. The answer graph of these nodes is initialized with the identifier (ids) of all the vertices that correspond to this label.
Next, we discover the differential constraints for each one of these initial lattice nodes. 
The discovery process for the differential constraints is described in detail in \Cref{subsec:horizontalExpansion}.

For each of the discovered set of differential constraints, we add another lattice node with the same graph pattern and the set of differential constraints in the lattice's same level (in this case, the first level).
Observe level $0$ in \Cref{fig:secondlevel_2}, which contains single-node patterns with and without differential constraints.
We create a new answer graph for each newly added lattice node in the horizontal expansion.
We filter the answer graph of the lattice node with the same graph pattern and empty set of constraints by removing nodes and edges that do not satisfy the set of differential constraints. Finally, we apply a node burn-back, a procedure that removes any disconnected nodes and edges, on this new answer graph (for more details on the node burn-back, see ~\cite{answergraph}).
For every lattice node $Q_0[\overline{x}]\phi_0$ whose support is larger than $\tau$, we add this node to a structure we call the candidate index used in the final step of GGDMiner, which details will be presented in \Cref{subsec:candidateindex}. 

To expand the lattice vertically, we use a frequent subgraph mining algorithm. At each vertical expansion, we add one edge to the graph pattern of the previous level lattice nodes.
In this implementation, we use the GRAMI~\cite{Elseidy2014} algorithm, which uses gSpan to generate candidates. 
At each edge extension (rightmost extension in gSpan) in which the frequency is higher than $\tau$, we add a new node to the lattice containing this new extended graph pattern $Q_1$ and an empty set $\phi$ of differential constraints.
We also create an answer graph for this new lattice node by extending the answer graph of the parent lattice node to $Q_1$ (in this case, $Q_0$) with the matches of the edge extended. 
At each edge extended in the answer graph, we run a node burn-back, which deletes from the answer graph all the disconnected nodes and edges that are no longer part of a match of this graph pattern. 

Next, we run a horizontal extension to discover differential constraints considering only the new variables included in the graph pattern using the new edge and new vertex that was added to the graph pattern (see an example of the first levels of the lattice in Figure \ref{fig:secondlevel_2}).
Similarly, for each set of differential constraints added, we add a lattice node in the same level and all these nodes in the Candidate Index. This described process of vertical expansions is executed recursively until the number of edges in the graph pattern of the lattice node is bigger than $k$. Algorithm \ref{algo:verticalExpansion} summarizes the vertical expansion of the lattice. 

\subsubsection{Discovery of Differential Constraints}
\label{subsec:horizontalExpansion}

The discovery of differential constraints relies on the attribute pairs selected in the pre-processing step of this framework. 
Given a graph pattern $Q[[\overline{x}]]$ in which $\overline{x}$ is the set of variables, we first verify which variables have attributes considered important according to the list in our pre-processing step. We discover differential constraints concerning only the variables (node/edge) that have attributes considered important. 
Thus, to avoid recalculations, we consider only the variables that were added to the graph pattern with the new edge extension at every vertical expansion of the lattice. Therefore, the number of possible differential constraints to be discovered at each extension is bounded.  
We take inspiration from methods for mining association rules with intervals to discover the differential constraints in GGDs. The main difference is that, in the case of differential constraints for GGDs, the intervals are thresholds, which indicates how similar two attributes or an attribute and a constant are. 

Given a graph pattern $Q[\overline{x}]$, its set of matches $E$, support $\tau$, and decision boundaries $T$, we first discover differential constraints of the type $\delta(x.A, cons) \le t$ where an attribute $A$ of a variable $x$ is compared to a constant $cons$. 
For each variable $x$ added to the graph pattern at the last extension, we first identify if there are any attributes of $x$ in the list of important attributes of the pre-processing step. If yes, we retrieve the set of values $V$ of attribute $A$ of $x$ in the set of matches $E$ and identify which clusters $C$ each value $v \in V$ is part of. 
Then for each identified cluster $c \in C$, if the size of $C$ is bigger than $\tau$, it means that there is at least one differential constraint regarding the attribute $x.A$. 
 
Considering the set of values of each cluster $c$, we choose the best value in the cluster $c$ to be used as the constant $cons$.  
The idea is to choose a constant value that maximizes the support of the differential constraint as much as possible, therefore, we select the value with the lowest average (dis)similarity to all other values in the cluster.
Finally, considering the value chosen for $cons$ and the (dis)similarity to the other value in the cluster, we identify which intervals of (dis)similarity values have support/frequency bigger than $\tau$. The intervals correspond to the threshold values of the discovered constraint. This function also considers the decision boundary of the attribute domain, which means that the discovered constraints should be at least $\upsilon$ similar to be considered a differential constraint, and the intervals discovered should have at least $\kappa$ difference between them. 
Any interval with higher support than $\tau$ is added as a new constraint.

We execute a similar algorithm to discover a differential constraint of the type $\delta(x.A,x'.B) \le t$, in which a variable attribute is compared to (possibly other) variable attribute. 
However, in this case, instead of verifying which attributes are on the list of most important attributes, we verify which attributes can be paired/compared to other attributes. Thus, since we do not have a cluster pre-built in the pre-processing step, instead, we use the pass join algorithm to cluster the values of each pair of attributes (same procedure as when building the similarity indexes in the pre-processing step) and repeat the process described before to discover the differential constraints.

\subsubsection{Candidate Index}
\label{subsec:candidateindex}

The candidate index is composed of two data structures: (1) a set $C$ of candidates that will serve as the source of the extracted GGDs and (2) an approximate $k$-NN graph~\cite{nndescent} in which each node represents a lattice node and the edges connect to the $k$ most similar nodes according to the similarity measure $\Delta$. 

The goal of this index is twofold:(1) to find the candidates that will be the source side of the extracted GGDs such that this set maximizes coverage and (2) to find target candidates that can be paired to the selected sources and minimize the verification of pairs of candidates. The target candidates are chosen according to how dissimilar they are to the source, the intuition behind this idea is that a very similar source and target might be trivial and not as interesting as a GGD in which the source and target have a certain degree of dissimilarity. 

We use a greedy-based approach to select the set of source candidates that maximizes the coverage. Whenever a new candidate $\{Q_l, \phi_l, E\}$ is added to the index (line 9 of Algorithm \ref{algo:verticalExpansion}), we verify if this newly added candidate can increase the coverage of the current set. If it can increase, we verify if there is a subset $S$ of $C$ in which $\{S\} \cup \{Q_l, \phi_l, E)\}$ has the same coverage as $\{C\} \cup \{(Q_l, \phi_l, E)\}$. If yes, then $\{S\} \cup \{Q_l, \phi_l, E)\}$ is the new set $C$, otherwise, $\{(Q_l, \phi_l, E)\}$ is added to $C$. The coverage of a set of GGDs is calculated using the candidate answer graph $E$, and there is no need to revisit the input graph.  

\subsection{GGD Extraction}

\begin{figure}[t]
    \centering
    \includegraphics[width=0.6\linewidth]{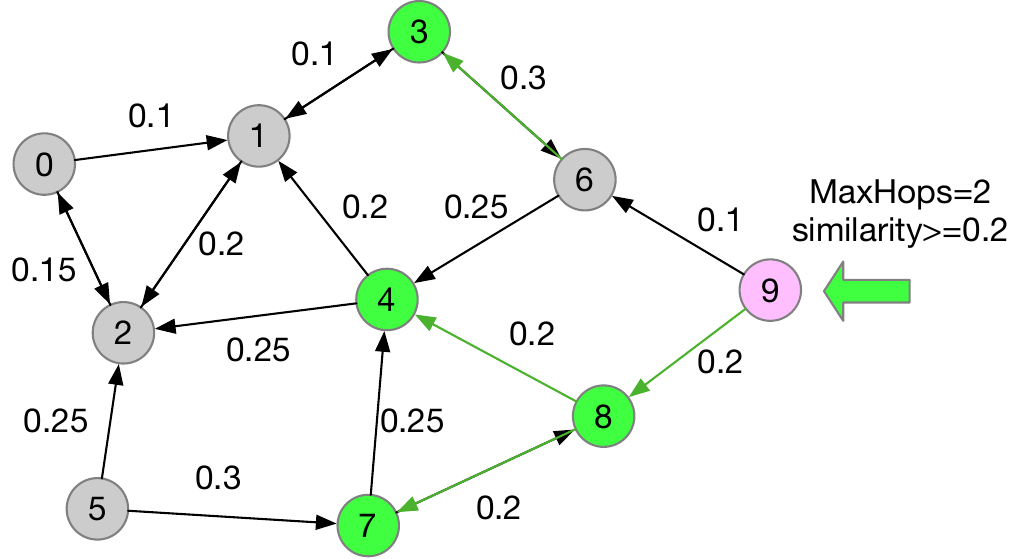}
    \caption{Extraction of possible lattice nodes which can possibly form a GGD with source lattice node 9}
    \label{fig:ggdExtractionResult}
    \end{figure}
\begin{figure}[t]
     \centering
    \includegraphics[width=\linewidth]{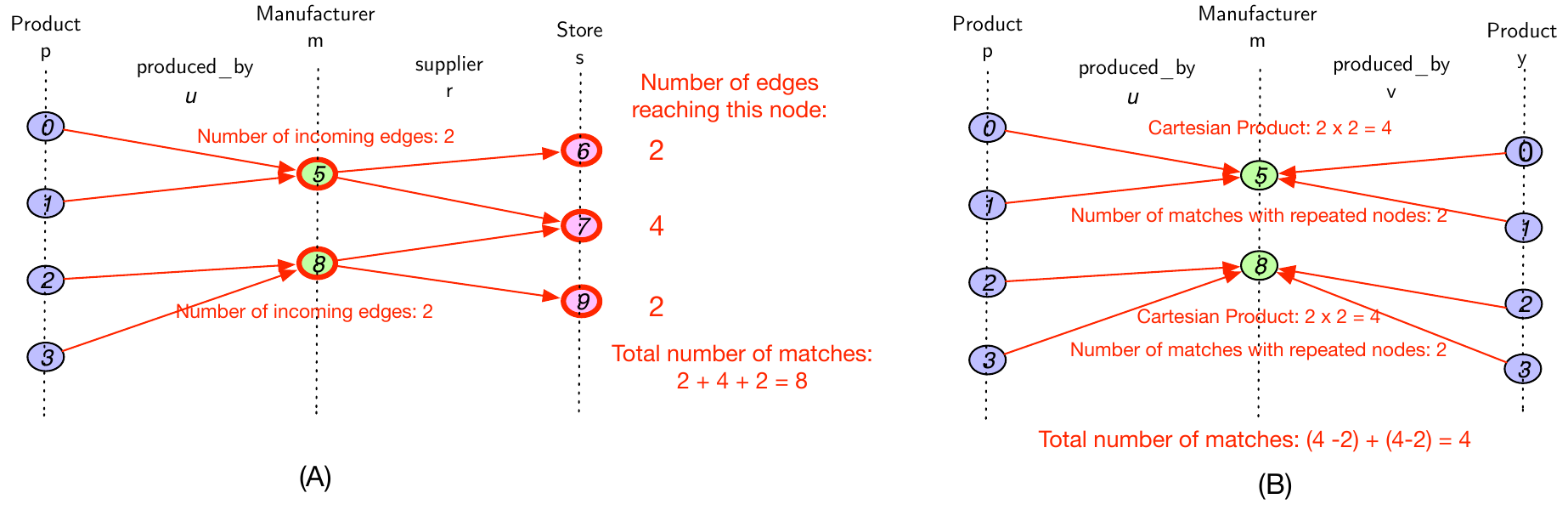}
    \caption{Calculating number of matches from Answer Graph for (A) a chain pattern and (B) a snowflake pattern}
    \label{fig:answer_graph_matches}
\end{figure}

The last step of the GGDMiner framework is the GGD Extraction. The main goal of this step is to extract the final resulting set of GGDs from the candidate index. In this step, we pair candidates of the candidate index as possible GGDs and verify which pairs have confidence value above the threshold $\epsilon$. 

We run a search algorithm on the $k$-NN graph to pair candidates that are not too similar to the point that is not interesting but still similar enough so that it is a valid GGD (confidence higher than $\epsilon$). 
Given a starting vertex $u$, a similarity threshold $\delta$, and a maximum number of hops $m$, we run a breadth-first search starting from vertex $u$ until it reaches $m$ hops from $u$. For each node $n$ visited during this search process, if $\Delta(u, n) \ge \delta$, we add $n$ to the result set $N$ of the search.  
In \Cref{fig:ggdExtractionResult}, we show the graph-based index of the Candidate Index with $10$ candidates from the candidate generation step and highlighted in green the possible target candidates of the source candidate with index $9$ (in pink).
We run this search starting from each vertices in the set $C$ in the candidate index. The set of pairs of the resulting $N$ set of the algorithm are the possible GGDs.
By using this search method, since we have a maximum number of hops from $u \in C$, regardless of the user-defined confidence value $\epsilon$, there is a maximum number of $k{_g}^m$ possible candidates to evaluate for each source candidate $u$, in which $k_g$ is the number of neighbors $k$ in the $k$-NN graph.

\begin{table}[t]
\small
\begin{tabular}{lll|ll|ll|l}
\cline{4-7}
 &  &  & \multicolumn{2}{l|}{Original} & \multicolumn{2}{l|}{GGDMiner} &  \\ \hline
\multicolumn{1}{|l|}{Dataset} & \multicolumn{1}{l|}{Nodes} & Edges & \multicolumn{1}{l|}{|N|} & |E| & \multicolumn{1}{l|}{|N|} & |E| & \multicolumn{1}{l|}{|$\Sigma$|} \\ \hline
\multicolumn{1}{|l|}{Cordis\tablefootnote{Graph built from Horizon 2020 project information accessed on \url{https://data.europa.eu/data/datasets/cordish2020projects?locale=en}}} & \multicolumn{1}{l|}{32K} & 151K & \multicolumn{1}{l|}{11} & 12 & \multicolumn{1}{l|}{8} & 8 & \multicolumn{1}{l|}{91} \\ \hline
\multicolumn{1}{|l|}{GDelt\tablefootnote{\url{https://github.com/smartdatalake/datasets/tree/master/gdelt}}} & \multicolumn{1}{l|}{73K} & 445K & \multicolumn{1}{l|}{5} & 2 & \multicolumn{1}{l|}{5} & 2 & \multicolumn{1}{l|}{61} \\ \hline
\multicolumn{1}{|l|}{DBLP\tablefootnote{\url{https://www.aminer.org/citation} }} & \multicolumn{1}{l|}{2M} & 810K & \multicolumn{1}{l|}{4} & 4 & \multicolumn{1}{l|}{4} & 3 & \multicolumn{1}{l|}{13} \\ \hline
\multicolumn{1}{|l|}{LDBC\tablefootnote{\url{https://github.com/ldbc/ldbc_snb_datagen_spark}}} & \multicolumn{1}{l|}{430k} & 2M & \multicolumn{1}{l|}{8} & 23 & \multicolumn{1}{l|}{7} & 20 & \multicolumn{1}{l|}{198} \\ \hline
\end{tabular}
\caption{Datasets and schema discovery results - |N| = number of node labels and |E| = number of edge labels. Parameters: $\tau=1000, \epsilon=0.7, k=2$, $|C|=7$ for DBLP and GDelt and $|C|=15$ for Cordis and LDBC}
\label{tab:datasets}
\end{table}

For each one of these candidate pairs in $N$, we verify if the confidence is above the threshold $\epsilon$.
Since we are looking for Extension GGDs, at least one common variable should exist in the source and target. We first verify the possible mappings we can have from the graph pattern from $c_1$ to $c_2$ in which at least one variable refers to the same nodes or edges in both graph patterns. If a mapping does not exist, this pair is discarded as a possible GGD.
To simplify the computation of the possible mappings, we use the idea of the DFS codes used by gSpan to enumerate the possible subgraphs common between both graph patterns. Then, we calculate the confidence of $c_1 \rightarrow c_2$ for each possible mapping. If $confidence(c_1 \rightarrow c_2) \ge \epsilon$, we rename the variables in $c_2$ according to our mapping and add $c_1 \rightarrow c_2$ as a GGD to our result set $\Sigma_G$. 

To calculate the confidence of a possible GGD, we calculate the source's total number of matches and the source's validated matches by using the Answer Graph of the source and target candidates. 
Considering the common variables between the source and target, we remove the matching nodes/edges that do not exist in the target Answer Graph from the source Answer Graph. Essentially, we filter the source Answer Graph to have only the matching nodes and edges that are validated by the target. Then, we calculate the number of matches that are represented by this filtered Answer Graph, which corresponds to the number of validated matches. 
The method for calculating the number of matches from the Answer Graph depends on the shape of the graph pattern. Observe in \Cref{fig:answer_graph_matches} how we calculate the number of matches without defactorizing the Answer Graph for these two types of shapes. This procedure allows to calculate the confidence of a GGD without defactorizing (extracting the matches) the source or the target Answer Graph in a table-like representation, thus, we do not need any extra information from the input graph $G$. 
By the end of this step, we have our final set of GGDs $\Sigma_G$. 

\section{Experimental Evaluation}
\label{sec:experiments}

In this section, we use the real-world datasets Cordis, GDelt, DBLP, and the synthetic dataset LDBC~\cite{ldbcData} generated with scale factor $0.1$ (see Table \ref{tab:datasets} for dataset size) to evaluate GGDMiner. We evaluate the impact of the main user-defined parameters on execution time and coverage of discovered GGDs and show examples of GGDs discovered by GGDMiner. 
Due to limited space, details about the datasets and implementation are available in the repository\footnote{\url{https://github.com/laricsh/ggdminer}}. 
We used $\epsilon=0.7$ and $\theta=0.5$ for all experiments unless mentioned otherwise. 
GGDMiner was implemented using Java and deployed on an Intel Xeon machine with 3.07GHz using 128GB of RAM.

\begin{figure*}[t]
   \centering
    \includegraphics[width=\linewidth]{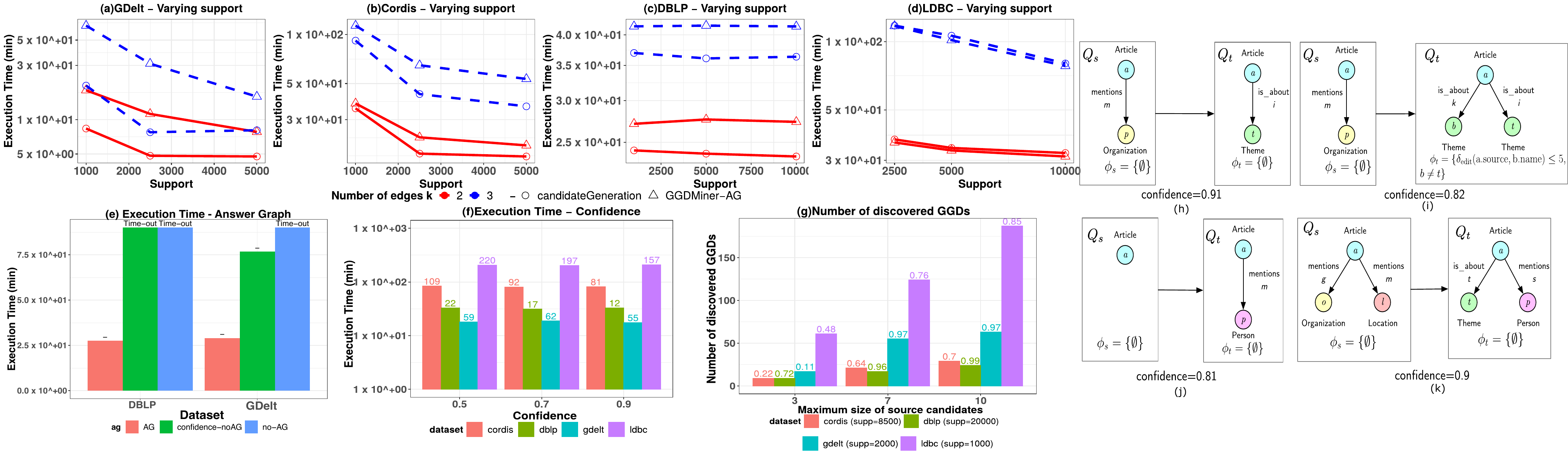}
    \caption{GGDMiner Experimental Results}
    \label{fig:resultsScalability}
\end{figure*}


\paragraph{Impact of discovery of graph patterns}
One of the main innovations of GGDs compared to previous data dependencies is how we can express a correlation between two (possibly different) graph patterns. In this first experiment, we evaluate the overhead in execution time of checking correlated graph patterns (GGD Extraction) compared to mining the graph patterns (Candidate Generation), a step also used in other data dependency algorithms. For this experiment, we run the GGDMiner without the differential set discovery to compare its behavior to that of frequent subgraph mining algorithms. 
We evaluate the execution time of GGDMiner according to the support $\tau$ parameter and also a maximum number of edges $k$ in the graph patterns, results are shown in \Cref{fig:resultsScalability}(a)-(d). These two parameters control the number of possible candidates of source and target the GGDMiner can have, as the smaller the value $\tau$ and bigger the $k$, the bigger the number of possible candidates of graph patterns and differential constraints to be evaluated. 

When comparing the execution time of GGDMiner to the execution of only Candidate Generation, we can observe that the execution time increases by approximately $1$ order magnitude for GDelt. The reason why is even though the number of candidates considered a source of a GGD is limited, and the number of candidates that can be considered target is also limited based on the candidate index, a pair of candidates can have multiple mappings (set common nodes or edges), and, for each mapping a confidence value must be calculated, increasing the execution time. The number of different mappings can be even higher when there is an entity/node label in the graph dataset that is the main subject of that dataset and appears in most graph patterns considered candidates. Since we considered datasets in the context of citation networks, this occurs in all datasets. Considering $\tau=1000$ and $k=2$, in the GDelt dataset, the FSM algorithm mined a total of $18$ graph patterns, which all include at least one node labeled ``Article''. In the DBLP dataset, all the mined graph patterns include a node labeled ``Paper'' of a total of $20$ mined graph pattern, and, in Cordis, from the total of $31$ mined graph patterns, $20$ of the mined graph patterns include a node labeled ``Project'' and $15$ include a node labeled ``Paper'', which means that for almost every pair of candidates considered as a possible GGD in the GGD Extraction step, there will be at least one mapping to check the confidence. However, the increase for Cordis and LDBC is less accentuated. This is because both datasets have a higher number of node and edge labels, and consequently, checking the possible frequent subgraphs takes more time, accentuating the overall execution time compared to GDelt and DBLP. 
We can also observe the difference in execution time according to parameter $k$ (maximum number of edges). 
In Cordis and in the LDBC dataset, given the bigger number of nodes and edge labels to be considered and evaluated as candidates in the candidate generation step, there is a bigger difference in execution time from $k=2$ to $k=3$ compared to other datasets. 

\paragraph{Impact of Answer Graph}
The Answer Graph is one of the key elements of GGDMiner. It is used to (1) represent the matched nodes and edges during the Candidate Generation step and (2) calculate the confidence value of GGDs without having to defactorize it (extracting the list of matches). 
We evaluate how much the Answer Graph improves the execution time in GGDMiner compared to a version of GGDMiner that uses a table-like representation of each graph pattern match for confidence checking.  The results are in \Cref{fig:resultsScalability}(e). Due to the long execution time of GGDMiner without the full use of Answer Graph, we fixed a maximum execution time of 1.5 hours. Results for Cordis and LDBC were not reported as the version that did not use Answer Graph ran out of memory.  

From the plots, we can easily identify how Answer Graph significantly improves the execution time of GGDMiner even for small graph patterns such as $k=2$ in several orders of magnitude. For both datasets, Cordis and GDelt, the execution time exceeded our fixed limit, while the execution time when using the Answer Graph was below the fixed limit. 
Not using Answer Graph for checking confidence also has the extra time overhead of defactorizing each source and target Answer Graph and verifying if each match of the source is validated on the target. 
On \Cref{fig:resultsScalability}(e), we also show the time difference between using the Answer Graph for calculating confidence compared to defactorizing it to a table-like representation of the matches (confidence-noAG). In this case, for GDelt, the execution time was less than half using Answer Graph compared to not using it, while for the other datasets, the execution time exceeded the fixed limit.
Calculating the confidence is related to validating GGDs, which has proven to be a high complexity problem in the literature~\cite{Shimomura2022}. Thus, this process can be repeated multiple times in GGDMiner as a pair of candidates might have multiple mappings (common source and target variables) that can form a GGD. 
While there is room for improvement in GGDMiner, Answer Graph has been shown to be a good alternative for representing the matched data.

\paragraph{GGDMiner vs. AMIE} We compared GGDMiner to AMIE+~\cite{Amie2015}. AMIE+ is an algorithm that mines Horn rules in knowledge bases. 
GGDs and rules mined by AMIE+ have different expressiveness, while AMIE+ mines rules in which the right-hand side is a single fact, GGDs can express a full graph pattern. 
In \Cref{tab:amie}, we compare the execution time for running AMIE+ and the number of output rules compared to GGDMiner for the LDBC and Cordis datasets for a number of $k$ edges (for GGDMiner) or facts (for AMIE+). Given the format of the rules of AMIE+ and the small schema of the graph, AMIE+ outputs an empty set of rules for GDelt and DBLP and, therefore, was not included in this comparison. 
Even though AMIE+ was faster than GGDMiner for the Cordis dataset, AMIE+ mined only one rule in comparison to the $67$ GGDs output by GGDMiner. 
In the LDBC dataset, AMIE+ could mine $31$, using a maximum of $2$ facts on the left-hand side of the rules and $79$ rules for a maximum of $3$. Nevertheless, the execution time for $k=3$ was double that of GGDMiner.  We also verified that $6$ of the rules mined by AMIE+ were included in the output set of GGDs. The small number of common rules/GGDs mined by the algorithms is due to the different expressiveness and heuristics used during each algorithm. While AMIE+'s goal is to find rules that can be used to infer a single fact, GGDMiner discovers GGDs that can cover most of the graph. 
We plan to extend GGDMiner using similar heuristics to AMIE+ and extend our evaluation in the future.

\begin{table}[t]
\small
\begin{tabular}{ll|l|l|l|l|l}
\cline{3-6}
                              &   & AMIE &         & GGDMiner &        &                               \\ \hline
\multicolumn{1}{|l|}{Dataset} & k & Time & |Rules| & Time     & |GGDs| & \multicolumn{1}{l|}{|common|} \\ \hline
\multicolumn{1}{|l|}{LDBC}    & 2 & 2    & 31      & 48       & 75     & \multicolumn{1}{l|}{6}        \\ \hline
\multicolumn{1}{|l|}{LDBC}    & 3 & 345  & 79      & 170      & 86     & \multicolumn{1}{l|}{6}        \\ \hline
\multicolumn{1}{|l|}{Cordis}  & 2 & 0.68 & 1       & 86       & 67     & \multicolumn{1}{l|}{0}        \\ \hline
\multicolumn{1}{|l|}{Cordis}  & 3 & 20   & 1       & 179      & 103    & \multicolumn{1}{l|}{0}        \\ \hline
\end{tabular}
\caption{GGDMiner vs. AMIE - Time (min), coverage (cov) and number of rules and GGDs mined by AMIE+ and GGDMiner}
\label{tab:amie}
\end{table}

\paragraph{Profiling Property Graphs}To demonstrate how the set of GGDs discovered by GGDMiner can cover underlying relations in a property graph, we evaluate the resulting set of GGDs according to coverage and number of output GGDs, apply it to schema discovery and present examples obtained.

\emph{Coverage and Number of output GGDs}
In this experiment, we evaluate the resulting set of GGDs from GGDMiner. 
First, we verify the execution time of GGDMiner according to the confidence value $\epsilon$ and the number of output GGDs, for these experiments, we fixed support value $\tau=1000$, $\epsilon=0.7$ and $k=2$ and $|C| = 7$. The results are shown in \autoref{fig:resultsScalability}(f). For the same support parameters, the execution time is similar for each dataset, independent of the confidence value. This is because Candidate Generation (which is sensitive to support) is the most time-consuming step for GGDMiner. Thus, for the same number of source candidates, independent of the confidence value, there is a maximum number of target candidates that each source will check. Nevertheless, confidence naturally affects the number of output GGDs, as the higher the confidence value, the more exact a GGD should be.

The size of the source candidates $|C|$ affects the number of output GGDs and the resulting coverage of the output set. As we can observe in \Cref{fig:resultsScalability}(g), by increasing the number of possible source candidates $|C|$, there is also an increase in coverage and the number of GGDs in the output set. However, we can observe that for all datasets, increasing the number of source candidates from $7$ to $10$ did not significantly increase in coverage compared to $3$ to $7$. Indicating that for each dataset, there is a candidate size value that will maximize coverage without many overlapping GGDs in the result set. 
Even with a small number of edges per pattern and a fixed size of source candidates, the resulting set from GGDMiner can still cover a big percentage of the input graph. Achieving over $90\%$ for DBLP and GDelt, $70\%$ on Cordis, and $85\%$ on LDBC.

\emph{Schema Discovery}  In this experiment, we use the discovered set of GGDs to rebuild the property graph schema. Since we have some preliminary information about the schema as input (labels, properties, and domain), we focus on identifying the relationships between the different node/edge labels in this experiment. To rebuild the schema using GGDs, we first execute GGDMiner without differential constraint discovery and discover a set $\Sigma$ of GGDs. Then, we build a schema graph that contains all the graph patterns found as source or target of the output set of GGDs and compare it to the original graph schema. 
The results of this comparison are available in \Cref{tab:datasets}. From the results, we can observe that even for datasets such as Cordis and LDBC, which have a higher number of node and edge labels, we were able to reconstruct a large part of the schema that agrees with the coverage obtained for each dataset.

\emph{GDelt Use Case}
The GDelt dataset comprises information from news articles sourced from the GDelt Project\footnote{\url{https://www.gdeltproject.org/}}, our study utilizes a subset of news articles\footnote{Subset information: \url{https://github.com/smartdatalake/datasets/tree/master/gdelt}}. 
For this study, we executed GGDMiner with $k=2, \epsilon=0.7, \tau=1000, \theta=0.5$
We identified a total of $61$ GGDs, achieving a coverage of $0.97$. In \Cref{fig:resultsScalability}(j)-(m), we display four of the obtained GGDs. Given that the confidence level of these GGDs is less than 1, they can highlight potential errors in the data or indicate unexpected relationships.

Although the graph has a limited number of nodes and edge labels, the set of discovered GGDs indicates that Article is the primary entity appearing in all discovered GGDs. Based on GGDs (h) and (i) shown in \Cref{fig:resultsScalability}, we observe the following: (1) Whenever an Article mentions an Organization, that Article is also linked to a Theme through the edge labeled ``IS\_ABOUT''; (2) GGD (i) further reveals that an Article can be associated with two distinct Theme nodes; (3) There appears to be a similarity between the “source” attribute of the Article and the “name” attribute, suggesting a relationship between the attributes of the nodes. If such GGDs are validated, violations might indicate the absence of an edge, potentially pointing to a missing connection to the Theme entity.
GGDs (j) and (k) illustrate the relationships between Article entities and both Location and Person. GGD (j) indicates that about 80\% (as suggested by the confidence value) of the articles in this dataset mention a Person. Meanwhile, GGD (k) demonstrates that an Article mentioning both a Location and an Organization also often references a Person and is linked to a Theme. Together, these GGDs highlight the connection of the various entities that an Article mentions, providing users with insights about the graph structure even when the full schema is not available. 
GGDs that aren't showcased in the figures and GGDs discovered from the Cordis, DBLP, and LDBC datasets can be found in our repository.

These analyses show the potential of the discovered GGDs to expose the input property graph's overall structure and more specific correlations between the graph patterns and their attributes. 

\section{Conclusion and Future Work}
\label{sec:conclusion}

In this paper, we proposed GGDMiner, a framework for discovering Graph Generating Dependencies (GGDs) for data profiling. GGDMiner can identify GGDs that succinctly describe a dataset. We employ the Answer Graph, a compact representation of the graph pattern matches, throughout the GGDMiner processes. This allows us to confirm which candidate pairs qualify as a GGD without the need for decompression (defactorization), improving the execution time of the framework.

GGDMiner is a framework-like solution for discovering GGDs with a common representation (Answer Graph), which can be used as a baseline for future contributions in the area. 
In future work, we plan to extend GGDMiner to support more methods in the implementation of each step of the framework and parallelize its execution to further improve scalability. 

\begin{acks}
Supported by the Deutsche Forschungsgemeinschaft (DFG, German Research Foundation) under Germany’s Excellence Strategy – EXC 2120/1 – 390831618 and by the European Union's Horizon 2020 research and innovation programme under grant agreement No 825041 and No 101058573.
\end{acks}


\bibliographystyle{ACM-Reference-Format}
\balance
\bibliography{sample-base}


\end{document}